# Model Predictive Control (MPC) Applied To Coupled Tank Liquid Level System

Med.Essahafi

*Abstract*— Coupled Tank system used for liquid level control is a model of plant that has usually been used in industries especially chemical process industries. Level control is also very important for mixing reactant process. This survey paper tries to presents in a systemic way an approach predictive control strategy for a system that is similar to the process and is represented by two liquid tanks. This system of coupled Tank is one of the most commonly available systems representing a coupled Multiple Input Multiple Output (MIMO) system. With 2 inputs and 2 outputs, it is the most primitive form of a coupled multivariable system. Therefor the basic concept of how the coupled tanks system works is by using a numerical system which it operates with a flow control valve FCV as main control of the level of liquid in one tank or both tanks. For this paper, MPC algorithm control is used which will be developed below. And it is focuses on the design and modelling for coupled tanks system. The steps followed for the design of the controller are: Developing a state space system model for the coupled tank system then design an MPC controller for the developed system model. And study the effect of the disturbance on measured level output. Note that the implementation Model Predictive Controller on flow controller valve in a Coupled Tank liquid level system is one of the new methods of controlling liquid level.

*Keywords*— Model predictive control, Level system, coupled Tank Plant..

## 1. INTRODUCTION

MODEL predictive control (MPC) has a long history in the field of control engineering. It is one of the few areas that has received on-going interest from researchers in both the industrial and academic communities.The general design objective of model predictive control is to compute a trajectory of a future manipulated variable input to optimize the future behavior of the plant output. The optimization is performed within a limited time window by giving plant information at the start of the time window.

Model predictive control, MPC, is a widely used industrial technique for advanced multivariable control. For processes with strong interaction between different signals MPC can offer substantial performance improvement compared with traditional single-input single-output control strategies.

Model predictive control has been used for several decades, and has been accepted as an important tool in many process industry applications.

The paper is organized as follows: Section 2 gives an overview of system modeling; section 3 describes briefly the underlying mathematics, for the Model predictive control algorithm using state space models. , section 4 focuses on simulation of the Model, and some conclusions are given in section 6.

TABLE I
LIST OF SYMBOLS

| Symbol | Quantity |
|---|---|
| $L1$ | Operating point Level in the tank 1 |
| $L2$ | Operating point Level in the tank 2 |
| $Fi1$ | Control flow rate into the tank 1 |
| $Fi1$ | Control flow rate into the tank 2 |
| $Fo1$ | Mutual leakage rate of flow of fluid between two coupled tanks. |
| $Fo2$ | Rate of flow of fluid from tanks 2 |
| $A1$ | Section area of the tank 1 |
| $A2$ | Section area of the tank 2 |
| $FCV1$ | A flow control valve of the tank 1 |
| $FCV2$ | A flow control valve of the tank 2 |
| $FIC$ | Regulator with flow indication, based on predictive control |
| $LT1$ | Transmitter level of the tank 1 |
| $LT2$ | Transmitter level of the tank 2 |
| $V1$ | Flow control valve 1 |
| $V2$ | Flow control valve 2 |
| $\alpha 1$ | Coefficient of discharge valve V1 |
| $\alpha 2$ | Coefficient of discharge valve V2 |
| $h1$ | Level manipulated Output variable in the tank 1 |
| $h2$ | Level manipulated Output variable in the tank 2 |
| $u1$ | Variable input rate of flow of valve 1 |
| $u2$ | Variable input rate of flow of valve 2 |
| $u3$ | Measured input Disturbance on rate of flow |
| fi1 | Flow manipulated Input variable in the tank 1 |
| fi2 | Flow manipulated Input variable in the tank 2 |
| Am | State matrix of state-space model |
| Bm | Input-to-state matrix of state-space model |
| Cm | State-to-output matrix of state-space model |
| Dm | Direct feed-through matrix of state-space model |
| ΔU | Parameter vector for the control sequence |
| Δu(ki + m) | Future incremental control at sample m |
| Ψ, Φ | Pair of matrices used in the prediction equation Y = Ψx(ki) + ΦΔU |
| Nc | Control horizon |
| Np | Prediction horizon |
| $o_m$ | Zero vector with appropriate dimension |
| x(ki + m \| ki | Predicted state variable vector at sample time m, given current state x(ki) |

M. Essahafi is with the Laboratory of Automation and Energy (LACE) Faculty of Science and Technology-CED-Science and Technology Conversion +212523-420-394 mohamed.essahafi@yahoo.fr

## 2. SYSTEM MODELING

The block diagram of the sample controlled two coupled tanks is shown in fig. 1 that comprises a two flow controller valve each of these valves is associated with its tank. The output real height level $h1, h2$ are measured and fed to controller FIC of flow inputs $u1, u2$ by level transmitter $LT1, LT2$. in the simulation studies described here, the nonlinear equations of two coupled tank plant are represented by a root square function.

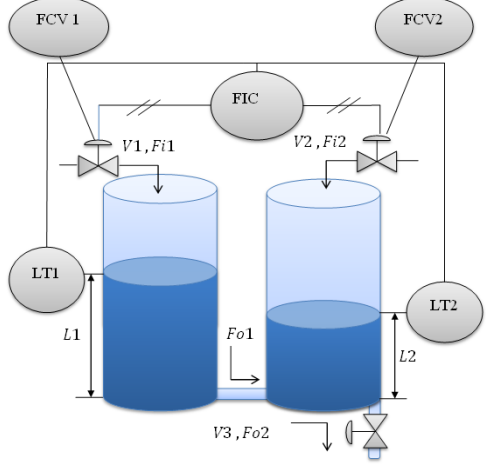

**Fig. 1** Level control sample of two coupled tanks

At any given time, the height of water level in either of the two tanks is associated to the water inlet rate, water outlet rate and the tank interactions.

$$A1.\frac{dL1}{dt} = Fi1 - Fo1 \qquad (1)$$

$$A2.\frac{dL2}{dt} = Fi2 - Fo2 + Fo1 \qquad (2)$$

Where, (3)

$$Fo1 = \alpha1.\sqrt{L1 - L2} \qquad (4)$$

$$Fo2 = \alpha2.\sqrt{L2} \qquad (5)$$

Replacing equations (4), (5) in equations (1) and (2), we obtain

$$A1.\frac{dL1}{dt} = Fi1 - \alpha1.\sqrt{L1 - L2} \qquad (6)$$

$$A2.\frac{dL2}{dt} = Fi2 - \alpha2.\sqrt{L2} + \alpha1.\sqrt{L1 - L2} \qquad (7)$$

The equations (6) and (7) represent a non-linear relationship between the water level (L1 and L2 in the two tanks, but if the operating point is known and does not change quite often then it is convenient to linearize the system obtained by first principles around the desired operating point. This makes the process significantly simpler and the model works well in a region around the chosen operating point. The stretch of operating band in which the linearized system gives a response similar to the actual nonlinear system determines the sensitivity of the linearized system.

To linearize the system around its operating point, a small change in flow input variables $fi1(u1)$ and $fi2(u2)$ is added which subsequently cause an incremental change in height in the two tanks $h1$ and $h2$ Hence, equations (6) and (7) can be rewritten as

$$A1.\frac{d(L1+h1)}{dt} = (Fi1 + fi1) - \alpha1.\sqrt{(L1+h1) - (L2+h2)} \qquad (8)$$

$$A2.\frac{d(L2+h2)}{dt} = (Fi2 + fi2) - \alpha2.\sqrt{(L2+h2)} + \alpha1.\sqrt{(L1+h1) - (L2+h2)} \qquad (9)$$

Subtracting equation (6) from (8) and (7) from (9) we have

$$A1.\frac{dh1}{dt} = fi1 - \alpha1.\left(\sqrt{(L1+h1) - (L2+h2)} - \sqrt{L1 - L2}\right) \qquad (10)$$

$$A2.\frac{dh2}{dt} = fi2 - \alpha2.\left(\sqrt{(L2+h2)} - \sqrt{L2}\right) + \alpha1.\left(\sqrt{(L1+h1) - (L2+h2)} - \sqrt{L1 - L2}\right) \qquad (11)$$

### 3. LINEAR MODEL OF PROCESS LEVEL FOR TWO COUPLED TANK SYSTEM

Once having developed the equations describing the system, MIMO linear model is needed to properly design the process control. Since the differential equations includes nonlinear terms. For this, Taylor expansion of the square root around a specific operating level $L1$ and $L2$ will be used to linearize the model. Once made a discrete time state representation will be defined for small $h1$ and $h2$ level variations around the operational level selected.

$$\begin{cases} x_m(k+1) = A_m x_m(k) + B_m u(k) \\ ym(k+1) = C_m x_m + D_m u(k) \end{cases} \qquad (12)$$

The Taylor expansion for $\sqrt{(1+x)}$ is given by:

$$(1+x)^n = 1 + n.x \quad ; \quad n=1/2 \qquad (13)$$

The equation (9) and (10) becomes:

$$\frac{dh1}{dt} = \frac{fi1}{A1} - \frac{\alpha1}{2.A1}\frac{1}{\sqrt{(L1-L2)}}.h1 + \frac{\alpha1}{2.A1}\frac{1}{\sqrt{(L1-L2)}}.h2 \qquad (14)$$

$$\frac{dh2}{dt} = \frac{fi2}{A2} + \frac{\alpha1}{2.A2}\frac{1}{\sqrt{(L1-L2)}}h1 + \frac{1}{2.A2}\left(\frac{-\alpha1}{\sqrt{(L1-L2)}} - \frac{\alpha2}{\sqrt{L2}}\right).h2 \qquad (15)$$

Where the matrices $Am, Bm, Cm$ and $Dm$ have the form:

$$A_m = \begin{pmatrix} -\frac{\alpha1}{2.A1}.\frac{1}{\sqrt{(L1-L2)}} & \frac{\alpha1}{2.A1}.\frac{1}{\sqrt{(L1-L2)}} \\ \frac{\alpha1}{2.A2}.\frac{1}{\sqrt{(L1-L2)}} & \frac{-1}{2.A2}.\left(\frac{\alpha1}{\sqrt{(L1-L2)}} + \frac{\alpha2}{\sqrt{L2}}\right) \end{pmatrix}$$

$$B_m = \begin{pmatrix} \frac{1}{A1} & 0 \\ 0 & \frac{1}{A2} \end{pmatrix} \quad C_m = \begin{pmatrix} 1 & 0 \\ 0 & 1 \end{pmatrix} \quad D_m = \begin{pmatrix} 0 & 0 \\ 0 & 0 \end{pmatrix}$$

Were,

$x = \begin{pmatrix} h1 \\ h2 \end{pmatrix}$   *State variable vector*

$u = \begin{pmatrix} fi1 \\ fi2 \end{pmatrix}$   *Control input vector manipulated*

$y = \begin{pmatrix} h1 \\ h2 \end{pmatrix}$   *Measurement vector*

The numerical values for parameters $L1, L2, A1$ $A2, A1$, and $A2$ are given in appendix.

## 4. MODEL PREDICTIVE CONTROL ALGORITHM

Model predictive control systems are designed by the mathematical model of the plant. The model to be used in the control system design is taken to be a state-space model. By using a state-space model, the current information required for predicting ahead is represented by the state variable at the current time.

The plant has 2 inputs (fluid flow rate *fi1* and fluid flow rate *fi2*), Also the number of outputs is 2 (height *h2* level and height *h2* level)

$$\begin{cases} x_m(k+1) = A_m x_m(k) + B_m u(k) \\ y_m(k+1) = C_m x_m + D_m u(k) \end{cases} \quad (16)$$

$xm$ Is the state variable vector with assumed dimension n1

$u(k)$ The input control. Thus, it is needed to change the model to suit the design purpose

However, due to the principle of receding horizon control, where current information of the plant is required for prediction and control, we have implicitly assumed that the input $u(k)$ cannot affect the output $y(k)$ at the same time. Thus,

$Dm = 0$ in the plant model.

Taking a difference operation on both sides of (equation 15), we obtain that

$$\begin{aligned} x_m(k+1) - x_m(k) &= A_m(x_m(k) - x_m(k-1)) \\ &+ B_m(u(k) - u(k-1)) \end{aligned} \quad (17)$$

Let us denote the difference of the state variable by

$$\Delta x_m(k+1) = x_m(k+1) - x_m(k) \quad (18)$$

And the difference of the control variable by

$$\Delta u(k) = u(k) - u(k-1) \quad (19)$$

These are the increments of the variables $x_m(k)$ and $u(k)$. With this transformation, the difference of the state-space equation is:

$$\Delta x_m(k+1) = A_m \Delta x_m(k) + B_m \Delta u(k) \quad (20)$$

Note that the input control to the state-space model is $u(k)$. the next step is to connect $\Delta x_m(k)$ to the output $y(k)$. to do so, a new state variable vector is chosen to be

$$x(k) = [\Delta x_m(k)^T \quad y(k)]^T \quad (21)$$

Where superscript T indicates matrix transpose. Note that

$$y(k+1) - y(k) = C_m(x_m(k+1) - x_m(k) \quad (22)$$

$$\begin{aligned} y(k+1) - y(k) &= C_m . A_m . \Delta x_m(k) \\ &+ C_m . B_m . \Delta u(k) \end{aligned} \quad (23)$$

Putting together (20) with (23) leads to the following state-space model:

$$\overbrace{\begin{bmatrix} \Delta x_m(k+1) \\ y(k+1) \end{bmatrix}}^{x(k+1)} = \overbrace{\begin{bmatrix} A_m & o_m^T \\ C_m A_m & 1 \end{bmatrix}}^{A} \overbrace{\begin{bmatrix} \Delta x_m(k) \\ y(k) \end{bmatrix}}^{x(k)} + \overbrace{\begin{bmatrix} B_m \\ C_m A_m \end{bmatrix}}^{B} \Delta u(k) \quad (24)$$

$$y(k) = \overbrace{[o_m \quad 1]}^{C} \begin{bmatrix} \Delta x_m(k) \\ y(k) \end{bmatrix} \quad (25)$$

Where

$$o_m = \overbrace{[0 \ 0 \ ..0]}^{n1} \quad (26)$$

The triplet $(A, B, C)$ is the augmented model, which will be used in the design of predictive control.

Here, we assume that the current time is $k_i$ and the length of the optimization window is $N_p$ as the number of samples. Assuming that at the sampling instant $k_i, k_i > 0$, the state variable vector $x(k_i)$ is available through measurement provided by the transmitter LT level shown in fig. 1, the state $x(k_i)$ provides the current plant information. The future control trajectory is denoted by

$$\Delta u(k_i), \Delta u(k_i+1), \ldots, \Delta u(k_i+N_c-1) \quad (27)$$

Where $N_c$ is called the control horizon dictating the number of parameters used to capture the future control trajectory. With given information $x(k_i)$, the future state variables are predicted for $Np$ number of samples, where $N_p$ is called the prediction horizon. $N_p$ is also the length of the optimization window. We denote the future state variables as

$$x(k_i+1|k_i), x(k_i+2|k_i) \ldots x(k_i+m|k_i), \\ x(k_i+N_p|k_i) \quad (28)$$

Where $x(k_i+1|k_i)$, is the predicted state variable at $k_i + m$ with given current plant information $x(k_i)$. The control horizon $N_c$ is chosen to be less than (or equal to) the prediction horizon $N_p$ based on the state-space model $(A, B, C)$, the future state variables are calculated sequentially using the set of future control parameters:

$$\begin{aligned} x(k_i+1|k_i) &= Ax(k_i) + B\Delta u(k_i) \\ x(k_i+2|k_i) &= Ax(k_i+1) + B\Delta u(ki+1) \\ &= A^2 x(k_i) + AB\Delta u(k_i) + B\Delta u(k_i+1) \end{aligned} \quad (29)$$

$$x(k_i + N_p|k_i) = A^{N_p}x(k_i) \\ + A^{N_p-1}B\Delta u(k_i) \ldots A^{N_p-N_c}B\Delta u(k_i \\ + N_c - 1) \quad (30)$$

From the predicted state variables, the predicted output variables are, by substitution

$$y(k_i + N_p|k_i) \\ = CA^{N_p}x(k_i) + CA^{N_p-1}B\Delta u(k_i) \ldots \\ + CA^{N_p-N_c}B\Delta u(k_i + N_c - 1) \quad (31)$$

Note that all predicted variables are formulated in terms of current state variable information $x(k_i)$ and the future control movement $\Delta u(k_i + j)$,
Where $j = 0, 1, \ldots N_c - 1$. Define vectors

$$Y = \begin{pmatrix} y(k_i + 1|k_i) \\ y(k_i + 2|k_i) \\ \vdots \\ y(k_i + N_p|k_i) \end{pmatrix} \quad (32)$$

$$\Delta U = \begin{pmatrix} \Delta u(k_i) \\ \Delta u(k_i + 1) \\ \vdots \\ \Delta u(k_i + N_c - 1) \end{pmatrix} \quad (33)$$

$$Y = \Psi x(k_i) + \Phi \Delta U \quad (34)$$

Where

$$\Psi = \begin{pmatrix} CA \\ CA^2 \\ \vdots \\ CA^{N_p} \end{pmatrix} \quad (35)$$

$$\Phi = \begin{pmatrix} CB & 0 & 0 & . & 0 \\ CAB & CB & 0 & . & 0 \\ CA^2B & CAB & CB & . & 0 \\ CA^{N_p-1}B & CA^{N_p-2}B & CA^{N_p-3}B & . & CA^{N_p-N_c}B \end{pmatrix} \quad (36)$$

## 5. OPTIMIZATION OF MPC

For a given set-point signal $r(ki)$ at sample time $ki$, within a prediction horizon the objective of the predictive control system is to bring the predicted output as close as possible to the set-point signal, where we assume that the set point signal remains constant in the optimization window. This objective is then translated into a design to find the 'best' control parameter vector $\Delta U$ such that an error function between the set-point and the predicted output is minimized.

Assuming that the data vector that contains the set-point information is

$$R_s^T = \overbrace{[1 \quad 1 \quad . \quad . \quad 1]}^{N_p} r(k_i) \quad (37)$$

We define the cost function J that reflects the control objective as

$$J = (R_s - Y)^T (R_s - Y) + \Delta U^T \bar{R} \Delta U \quad (38)$$

To find the optimal $\Delta U$ that will minimize J, by using the equation (38), J is expressed as

$$J = (R_s - \Psi x(ki))^T(Rs - \Psi x(ki)) - 2\Delta U^T \Phi^T (R_s - \Psi x(ki)) + \Delta U^T (\Phi^T \Phi + \bar{R})\Delta U. \quad (39)$$

From the first derivative of the cost function J:

$$\partial J / \partial \Delta U = -2\Phi^T (R_s - \Psi x(ki)) + 2(\Phi^T \Phi + \bar{R})\Delta U \quad (40)$$

The necessary condition of the minimum J is obtained as

$$\partial J / \partial \Delta U = 0 \quad (41)$$

From which we find the optimal solution for the control signal as

$$\Delta U = (\Phi^T \Phi + \bar{R})^{-1} \Phi^T (R_s - \Psi x(ki)) \quad (42)$$

With the assumption that $(\Phi^T \Phi + \bar{R})^{-1}$ exists. The matrix $(\Phi^T \Phi + \bar{R})^{-1}$ is called the Hessian matrix in the optimization literature. Note that $R_s$ is a data vector that contains the set-point information expressed as

$$R_s = \overbrace{[1 \quad 1 \quad . \quad . \quad 1]}^{N_p} r(k_i) = \overline{R_s} r(k_i) \quad (43)$$

The optimal solution of the control signal is linked to the set-point signal $r(k_i)$ and the state variable $x(k_i)$ via the following equation:

$$\Delta U = (\Phi^T \Phi + \bar{R})^{-1} \Phi^T (\overline{R_s} r(k_i) - \Psi x(ki)) \quad (44)$$

## 6. SIMULATION AND DISCUSSION

In Matlab for example, the first step is to create the augmented model for MPC design. And the input parameters to the function are the state-space model $(Am, Bm, Cm)$, prediction horizon $N_p$ and control horizon $N_c$. Then calculate the $\Psi$ and $\Phi$ matrices. And Calculate $\Delta U$ by assuming the information of initial condition on $x$ and $r$.

The Initial conditions for the process nonlinear system operating level are $L1 = 4\,m$ and $L2 = 3.5m$

The process model is as follow

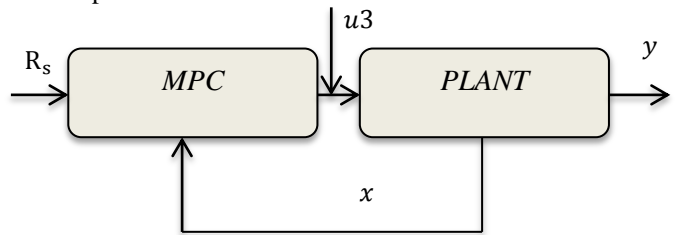

**Fig. 2** Simulation model used in closed loop MPC

$u3$ A disturbance on input rate of flow of fluid expressed in percentage of 100%.

The simulation has been done with respect to the following considerations:

The MPC Model has two references set point as input values, and two output flow control $u1, u2$ and a one disturbance on the control input $u3$.

The simulation has been done with respect to the following considerations:
Parameters of *MPC* controller

Prediction horizon $N_p = 10$

Control horizon $N_c = 3$

Sampling time 0.05

The reference is chosen as a pulse signal with size $h1 = 0.5m$ and $h2 = 0.3m$
The disturbance is chosen as a pulse signal with size $u3 = 10\%$.

In the fig.3 and fig.4 above, it can be observed that the control inputs variables flow $fi1$ and flow $fi2$ increase from the operating rate of flow $Fi1$ and $Fi2$ respectively in the same track of the variation of the measurement output variable $h1$ and $h2$. so in this case the MPC controller is in a direct sense.

The following case study illustrates best tracking and robustness with no oscillation and the ability of the proposed MPC to robustly maintaining best dynamic performance. The two coupled tanks discussed above is to be controlled by the proposed robust MPC found in equations (44). The case studies assume that no effect from the onset of disturbances is affecting the plant.

Fig. 3 below shows the performance of the unconstrained system responding to a pulse set point change.

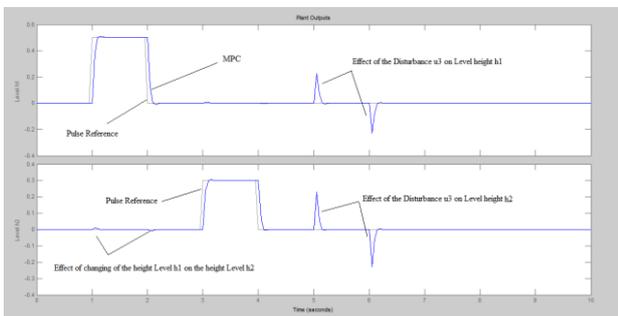

**Fig. 3** Measured output level $h1, h2$

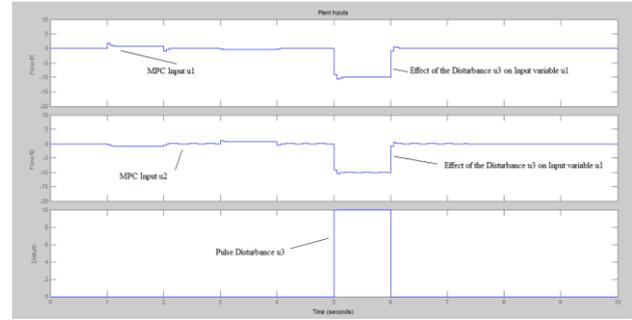

**Fig. 4** Manipulated inputs variables $u1, u2$ and $u3$
For the plant we have,

$$A_m = \begin{pmatrix} -7.923 & 7.923 \\ 9.781 & -12.97 \end{pmatrix} \quad B_m = \begin{pmatrix} 5.093 & 0 \\ 0 & 6.288 \end{pmatrix}$$

$$C_m = \begin{pmatrix} 1 & 0 \\ 0 & 1 \end{pmatrix} \quad D_m = \begin{pmatrix} 0 & 0 \\ 0 & 0 \end{pmatrix}$$

## 7. CONCLUSION

In this paper, a Model predictive controller is designed for a sample two coupled tanks comprising many tools of control. From the simulation results, it is clear that the MPC control is very suitable for nonlinear processes. Therefor the MPC controller allows for basic a good disturbance rejection and good robustness to model errors. Thus, we can design other models of process control level as the method of generalized predictive control GPC which may include disturbances and noise on the inputs and outputs.

## 8. APPENDIX

Numerical values for parameters $L1, \alpha1, \alpha2, L1, L2, A1, A2$.

$\alpha1 = 2,2;$
$\alpha2 = 1,9;$
$L1 = 4m;$
$L2 = 3,5m;$
$A1 = 0.1963 \; m^2$
$A2 = 0.159 \; m^2$